# QUALITY OF A QUANTUM ERROR CORRECTING SCHEME AND MEMORY ERROR THRESHOLD ESTIMATION

PEDRO J. SALAS

*Universidad Politécnica de Madrid*
psalas@etsit.upm.es

The error correcting capabilities of the Calderbank-Shor-Steane [[7,1,3]] quantum code, together with a fault-tolerant syndrome extraction by means of several ancilla states, have been numerically studied. A simple probability expression to characterize the code ability for correcting an encoded qubit has been considered. This probability, as a correction quality criterion, permits the error correction capabilities among different recovery schemes to be compared. The memory error threshold is calculated by means of the best method of those considered.

## 1   Introduction

Noise is a ubiquitous and damaging problem to be solved before building a quantum computer in the near future. A possible strategy for controlling the error infection of the states involved in a quantum computer is the use of a quantum error correcting code (QECC) [1]. How to use these QECCs in order to control the state decoherence coming from the interaction with the environment or from the faulty gate application is well known. Even though the methods used for constructing a quantum error correcting fault-tolerant network have already been established [2], its analysis is not straightforward, so a numerical simulation could provide valuable conclusions.

The aim of this work is to study the quantum error correction ability to control the decoherence throughout the time, when the recovery process includes different ancilla states. Usually the fidelity has been used to measure the correction capabilities of the code. But in this paper it has been split to demonstrate that the correctable and non-correctable error probabilities are more appropriate to analyze the error correction ability. A numerical simulation is carried out, introducing the noise by means of the depolarizing error model and using the [[7,1,3]] CSS code. The results will be compared with those of a simple non fault-tolerant error correction and will provide a threshold for memory errors with the best ancilla state. Keeping the main hypothesis, the results can be extrapolated to other quantum error correcting codes.

## 2   Error model

Quantum errors are analogical and their correction will require the following stages. The first step in correcting an error is to digitalize it into one of the following error operators [3]: I (identity, not being a strict error), X (bit-flip error), Z (phase-flip error) and Y (bit and phase flip error). This can be achieved because the one qubit Pauli error set is a basis for any 2×2 complex matrix. Secondly, to extract the error syndrome, a high fidelity ancilla state has to be previously prepared, and the error syndrome is copied onto the ancilla. Subsequently, the ancilla state is measured destructively in order to find out the error, allowing correction of the error state.

Quantum fidelity is defined as the overlap between a reference state and the current one in a particular noisy channel. The fidelity of fault-tolerant error correction depends heavily on the ability to synthesize high quality ancilla states. So it would be interesting to check how different ancilla states affect the ability to control and correct errors when a specific error model for the noisy quantum channel and quantum encoding are considered.

In the present work, the CSS [[7,1,3]] quantum code [4] using different ancilla states, is studied from the point of view of its effectiveness in controlling the qubit decoherence. A logical qubit (information qubit, IQ) is encoded by means of a noisy and non-fault-tolerant quantum network (process taken as a reference noisy computation) and sent through a quantum channel whose noise is introduced by means of the isotropic depolarizing error model [6]. To simulate the behaviour of quantum networks we use an *independent stochastic error model* based on the notion of *error locations* [5]. At each network location the error is applied at random and independently of other errors in the same or different locations. The evolution error (coming from the free evolution time steps in the quantum network) is introduced by means of the X, Y and Z errors in each qubit and time step, having the same error probability $\varepsilon/3$, with no error (I unity operator) appearing with a probability $(1-\varepsilon)$. Noisy one-qubit gates (Hadamard and measurement), have $\gamma$ error probability at each gate location. In the two-qubit gates (CNOT) case [6], we assume there are sixteen possibilities corresponding to the tensor product {I, X, Y, Z} ⊗ {I, X, Y, Z}. If the one-qubit gate error probability is $\gamma$, each two-qubit gate error appears with probability $\gamma/15$, because the I⊗I term is not, actually, an error operation. To introduce the noise, we let the gate operate before the error is introduced. This $O(\gamma)$ (instead of $O(\gamma^2)$) two-qubit error behaviour clearly over-estimates the difficulty of error correction, although it is not an unrealistic assumption.

Five different ancilla states have been used; four of them including a verification piece of network to check the state before to letting it interact with the IQ. Ancilla network construction and error model assumptions are detailed in [7] and will not be reproduced here. The ancillas are named with the same notation: (1)

is a simple three-qubit ancilla state without verification; (2) five-qubit Shor's ancilla; (3) Steane's ancilla; (4) parallelized Steane's ancilla; (5) parallelized Steane's ancilla with bit and phase-flip error verification network. Except in the case of ancilla (1), in which only one syndrome is obtained, three syndromes are calculated when ancillas (2) to (5) are used, taking the most repeated one to correct the IQ state. No action is carried out if the three syndromes are different in order to not worsen the IQ state.

In [7] we concluded that a simple ancilla state (1) provides reasonable results, even if it is not fault-tolerant. In fact, when only one syndrome was obtained, this ancilla state provides fidelity results comparable (even slightly better) than those reached by means of a fault-tolerant recovery network (4) or (5). We guessed that this result could come from the fact that only one recovery step was carried out. Considering a correction through the time, ancilla (1) would show less capability to control the decoherence than the remaining ancilla states. In addition in [7] the fidelity was used to quantify the quality of the corrected logical qubit. Unfortunately this quantity does not completely reflect the origin of its decreasing values. We calculate the infidelity (1-Fidelity) as representing the error probability and, whereas errors of weight one, two and three participate in diminishing this value, not all of them are equally harmful. Errors of weight one are not so bad, because they could be corrected at a later time step or after a measurement, considering the resultant codeword as a classic one and correcting it. Errors of weight two and three are not recoverable so they will accumulate in the IQ state, thus increasing the decoherence.

## 3 Quality of the error correction

A new error correcting quality tool for quantifying the ability to recover the IQ state is needed. This measure should distinguish between the correctable and non-correctable errors that contribute to diminishing the fidelity. The starting point is a binary QECC with codewords of length n, defined as the subspace $[|0_L\rangle, |1_L\rangle]$, generated by two encoded or logical qubits, having distance $d = 2t+1$ and correcting t errors. The full Hilbert space could be covered in the following way: consider error operators $E_u = A_{u_1} \otimes \ldots \otimes A_{u_n}$ with $A_{u_i} \in \{I, X, Y, Z\}$ and $u = (u_1,\ldots,u_n) \in$ GF(2)$^n$ (Galois field of binary vectors of length n) defining which physical qubits are in error. Considering errors with weight $W(u) = 0,..,t$ (the summation is taken for those u having weight $0 \leq W(u) \leq t$):

$$H^{(n)} = \bigoplus_{W(u)=0}^{t} E_u [|0_L\rangle, |1_L\rangle] \qquad (1)$$

From the point of view of error correction, and assuming perfect decoding, errors with weight $W(u) = 0,..,t$, are correctable. After a quantum computation process is carried out to recognize the result, a final measurement of the qubit state is taken to get a quantum register. If it is affected by t errors, at most, after a classical (faultless) error recovery the correct qubit identification is achieved assuming perfect decoding.

In spite of using the fidelity as the quality criterion, the following is proposed. Firstly a reference initial encoded state is chosen $|\phi_{ref}\rangle$ and sent through the quantum correcting network. After the correction the state $|\psi_f\rangle$ is obtained and the probability $P_C$ of $|\psi_f\rangle$ having a correctable error is:

$$P_C = |\langle\phi_{ref}|\psi_f\rangle|^2 + \sum_{W(u)=1}^{t}|\langle\phi_{ref}|E_u|\psi_f\rangle|^2 = F_0 + F_t \qquad (2)$$

$F_0$ being the usual fidelity and $F_t$ the fidelity after the correction of the errors with weight from 1 to t. The non-correctable error probability, coming from errors of weight $W(u) > t$, is $P_{NC} = 1-P_C$. $P_C$ means the probability that the state after error correction is found inside a sphere centred in the $|\phi_{ref}\rangle$ state and with radius $t = (d-1)/2$. In the case of a t-QECC having a distance $d > 2t+1$, the full Hilbert space is not completely covered by the direct sum of the subspaces translated by error operators of weight up to t (equation (1)), but the $P_C$ equation (2) is still valid to quantify the correction quality.

As $P_C$ depends on the QECC used, the recovery network, the reference state and particularly the error model assumed, their optimal value will be difficult to obtain in general. Fortunately, an estimation can be reached by taking one of the code basis $|\phi_{ref}\rangle = |0_L\rangle$ or $|1_L\rangle$ as the reference state and using a concrete error model. This dependence reflects that $P_{NC}$ does not provide an absolute number characterizing the code quality but permits the comparison between several methods or codes to distinguish the good one.

In the following, this criterion is applied to the [[7,1,3]] quantum correcting code using several ancilla states. This code is a CSS (Calderbank-Shor-Steane) with distance three [4], so it corrects one bit and phase-flip independently, and the error operators of weight one can be expressed as $E_u = X_v Z_w$, $u, w \in GF(2)^n$, with $W(u) = W(v) = W(w) = 1$. If two errors (bit and phase-flip) happen in the same physical qubit, $E_u$ represents a $Y_u$ error, so it is also correctable. The total Hilbert space has dimension $\dim(H^{(7)}) = 128$, distributed in 64 subspaces of dimension two, characterized as having an orthonormal logical basis states $X_v Z_w [|0_L\rangle, |1_L\rangle]$, with v, w representing the physical qubit affected by the X or Z error operator. Note that the full Hilbert space is only covered with error operators of weight one:

$$H^{(7)} = \left[|0_L\rangle, |1_L\rangle\right] \oplus \left(\bigoplus_{W(v),W(w)=1}^{7} X_v Z_w \left[|0_L\rangle, |1_L\rangle\right]\right) = \bigoplus_{W(i),W(k)=0}^{7} X_i Z_k \left[|0_L\rangle, |1_L\rangle\right] \quad (3)$$

The value i = k = 0 representing the code subspace. The non-correctable error probability is now:

$$P_{NC}(\varepsilon,\gamma,t) = 1 - \sum_{\substack{W(i)=0 \\ W(k)=0}}^{1} |\langle 0_L | X_i Z_k | \psi_f \rangle|^2 = 1 - F_0(\varepsilon,\gamma,t) - F_1(\varepsilon,\gamma,t) \qquad (4)$$

Errors of weight two are non-correctable and those of weight three are also non-correctable because they are non-detectable. Both types contribute to $P_{NC}$, its value telling us how good the error correction is. Note that errors of weight greater than three are equivalent to another one with weight between one and three [6].

## 4  Simulations results

The ability for different ancilla states to control and correct errors has been considered in two cases: encoding with and without noise. Errors (X, Y, Z) are introduced into the calculation using the Luxury Pseudorandom Numbers [8] which is an improvement on the subtract-and-borrow random number generator proposed by Marsaglia and Zaman. The fortran-77 code is as a result of James [9], and is used with the luxury level parameter p = 389: the highest possible luxury value. The code returns a 32-bit random floating point number in the range (0, 1). For each run a new random seed is chosen as a 32-bit integer.

The simulation results start with the noisy encoding of the physical qubit |0> (taken as reference state) and then sends it through a noisy depolarizing channel. The error correction is carried out as frequently as possible, including only one time step ($\Delta t = 1$) between two consecutive recoveries.

For each time step the fidelities $F_0$, $F_1$ and the $P_{NC}$ probability have been calculated for the typical values $\varepsilon = 0.0002$ and $\gamma = 0.001$. Figure 1 shows the usual fidelity $F_0(\varepsilon=0.0002, \gamma=0.001, t)$ for the five different ancilla states considered.

Not all errors of weight one are eliminated because some of them appear at the end of the network so, as the numerical simulation shows the fidelities $F_0(\varepsilon,\gamma,t)$ are perfectly fitted to the straight lines of a negative slope $F_0(\varepsilon,\gamma,t) = - A(\varepsilon,\gamma)\, t + B(\varepsilon,\gamma)$. The initial values of $F_0$, after the first error correction, reflect the same relative recovery quality obtained in [7]. Over time, the slope of the fault-tolerant recovery strategies remain, approximately the same. The slope of $F_0(\varepsilon=0.0002, \gamma=0.001, t)$ for the simple ancilla (not fault-tolerant), reveal a clear inability to control the error accumulation in the encoded qubit $|0_L\rangle$. It is possible to distinguish the weight of the errors contributing to diminishing the fidelity value. $F_1$ is the fidelity coming from errors of weight one and is plotted versus the time steps in figure 2. Its value is constant for the case considered, so the different slopes of $F_0(\varepsilon,\gamma,t)$ account for the errors of weights two and three.

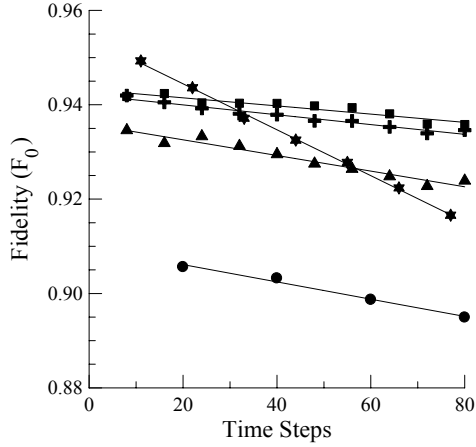

Figure 1. Fidelity $F_0$ against time steps for the noisy $|0_L\rangle$ qubit sent through a noisy depolarizing quantum channel with errors $\varepsilon = 0.0002$, $\gamma = 0.001$. The recovery uses the following ancilla states: ✶ simple ancilla, • Shor, ▲ Steane, ■ Steane parallelized and ✚ Steane parallelized with bit and phase-flip error verification.

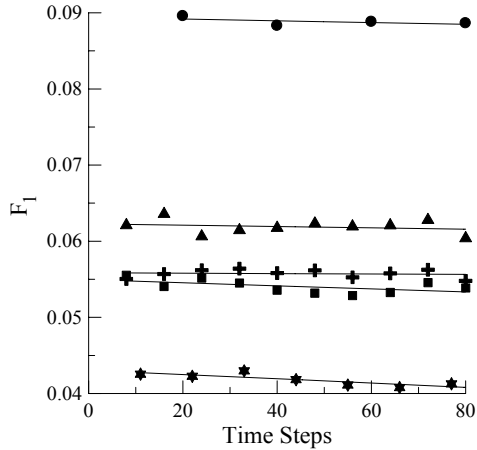

Figure 2. Fidelity $F_1$ against time steps for the noisy $|0_L\rangle$ qubit sent through a noisy depolarizing quantum channel with errors $\varepsilon = 0.0002$, $\gamma = 0.001$. Symbols are the same as in figure 1

By means of the criterion established in equation (4), the non-correctable error probability $P_{NC}(\varepsilon=0.0002, \gamma=0.001, t)$ has been calculated. The results are plotted in figure 3. There is a clear difference between the recovery using the non-fault-tolerant simple ancilla state and the remaining ancillas. For the values of $\varepsilon$ and $\gamma$ considered, ancilla (1) is clearly the worst, next Shor (2) and followed by Steane (3). Ancillas (2) and (3) have similar $P_{NC}$ probability, (3) being more suitable because of its smaller $F_1$ value. According to the quality criterion used, the best ancillas are (4) and (5), providing similar results for $P_{NC}(\varepsilon=0.0002, \gamma=0.001, t)$ and $F_1$. As a consequence, the inclusion of an additional phase error verification step in the ancilla (5) is not clearly beneficial to increasing the code capability to control the error accumulation, at least in the way it has been introduced. So the simplest and effective ancilla is the Steane parallelized ancilla (4). Figure 3 also includes the error probability when no encoding is used $P_{NE}(\varepsilon,t) = 1-(1-2\varepsilon/3)^t$, with $\varepsilon = 0.0002$. This expression comes from considering the non-encoded $|0\rangle$ qubit corrupted by two (X and Y) of the three possible errors. For the $\varepsilon$ and $\gamma$ considered, the encoded $|0_L\rangle$ qubit becomes more stabilized over time (smaller slope) when it is encoded and corrected by means of the ancillas (4) and (5).

The noisy encoding network is not fault-tolerant so neither is the whole process. In order to make the fault-tolerant behaviour more evident, a noise-free encoding has been considered. The perfect $|0_L\rangle$ is sent through a noisy depolarizing channel (with $\varepsilon$ evolution error) and periodically corrected with one time step ($\Delta t=1$) between consecutive corrections. The ancilla state is synthesized by means of a noisy network affected by $\varepsilon$ evolution error and $O(\gamma)$ gate error according to the previous model.

The non-correctable error probability originates in the errors of weights two and three in the case of the present code, because it has distance three. If the recovering method is fault-tolerant, keeping the relationship $C = \varepsilon/\gamma = 0.5$ constant and by varying $\varepsilon$, it must fulfil $P_{NC}(\varepsilon, C,t) \sim B\varepsilon^2$. The simulation results are shown in figure 4 and reflect the expected quadratic law. The values of B for the different ancillas are the following: $B_2 = 23990.422$, $B_3 = 12198.812$ and $B_4 = 5729.543$, the subscripts refer to the method used.

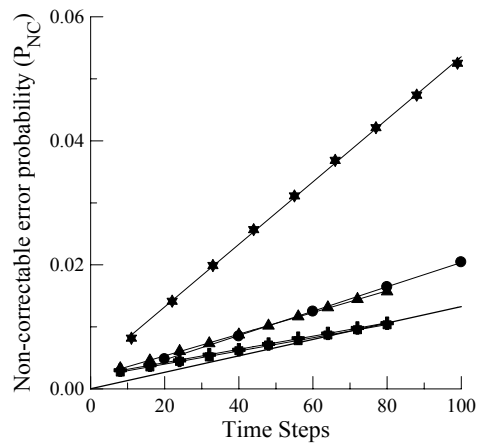

Figure 3. Non-correctable error probability $P_{NC} = 1-F_0-F_1$ against time steps for the noisy $|0_L\rangle$ qubit sent through a noisy depolarizing quantum channel with errors $\varepsilon = 0.0002$, $\gamma = 0.001$. Symbols are the same as in figure 1. Continuous line without symbols represent the error probability when no encoding is used, $P_{NE} = 1-(1-2\varepsilon/3)^t$, with $\varepsilon = 0.0002$.

Clearly the non-fault-tolerant simple ancilla has a linear behaviour $P_{NC}(\varepsilon, C, t) \sim O(\varepsilon)$, so it is less useful when $\varepsilon \to 0$. The rest of the ancilla states confirm a fault-tolerant behaviour, having the ancillas (4) and (5) coincident $P_{NC}(\varepsilon, C, t)$ results. Figure 4 does not include the results for ancilla (5) because they are very close to those of ancilla (4). The merit of ancilla (4) is having a simpler implementation circuit, so it is concluded that it is the best of those considered.

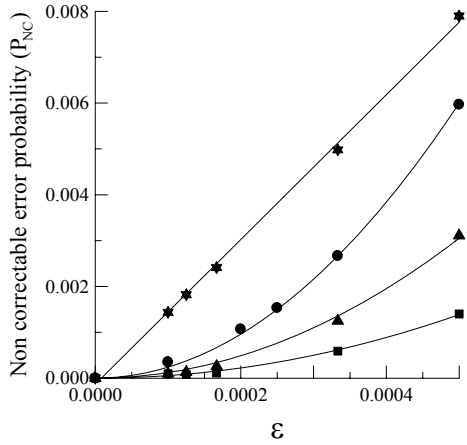

Figure 4. Non-correctable error probability ($P_{NC}$) against $\varepsilon$, (keeping constant the relationship $C=\varepsilon/\gamma=0.5$) for a perfect $|0_L\rangle$ qubit sent through a noisy depolarizing quantum channel affected by $\varepsilon$ evolution error and one time corrected by means of a recovering method using different noisy ancillas. Symbols are the same as in figure 1.

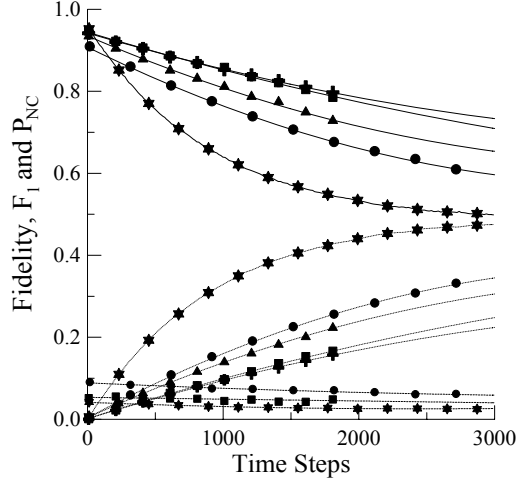

Figure 5. Fidelity and error probabilities against time steps for ε=0.0002 and γ=0.001. Upper set of lines: fidelity. Medium set: non-correctable error probability ($P_{NC}$). Bottom set: one error fidelity ($F_1$). The recovery uses the following ancilla states: ✶ simple ancilla, • Shor, ▲ Steane, ■ Steane parallelized and ✚ Steane parallelized with bit and phase-flip error verification. In the bottom set the three Steane curves considered are coincident, so only Steane parallelized is included (■).

Figure 5 shows the large time behaviour of the fidelity $F_0$(ε=0.0002, γ=0.001, t≤2000), $F_1$(ε=0.0002, γ=0.001, t≤2000) and $P_{NC}$(ε=0.0002, γ=0.001, t≤2000) for different ancillas. Note the value of $C = ε/γ = 0.2$ is located in the suitable correction region of figure 9, as will be explained next. For one time step the non-fault-tolerant ancilla (1) network provides quite good results. The behaviour changes dramatically as the time increases, having ancilla (1) an important error accumulation as $P_{NC}$ shows. While fidelity $F_1$(ε=0.0002, γ=0.001, t≤2000) seems not be determinant in the qubit quality, an important difference can be appreciated between ancilla (1) and the remaining. When the recovery includes one time step (Δt=1) between consecutive corrections, fault-tolerant ancillas are much more efficient. Some of the results have been calculated only until t ≈ 2000 and then extrapolated until t = 3000 steps.

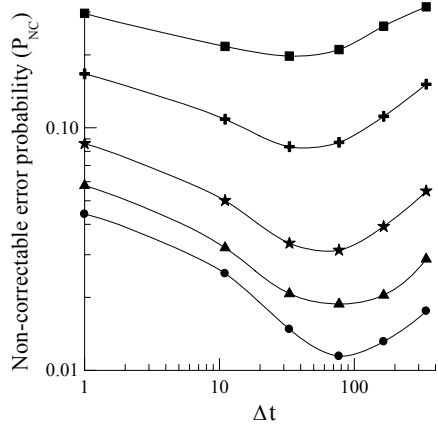

Figure 6. Non-correctable error probability ($P_{NC}$) against the number of time steps between consecutive corrections ($\Delta t$) using non-fault-tolerant ancilla (1). Maintaining C=0.9, the values of $\varepsilon$ are: ● $1.25\ 10^{-4}$, ▲ $1.66\ 10^{-4}$, ✶ $2.5\ 10^{-4}$, ✚ $5\ 10^{-4}$, ■ $10^{-3}$.

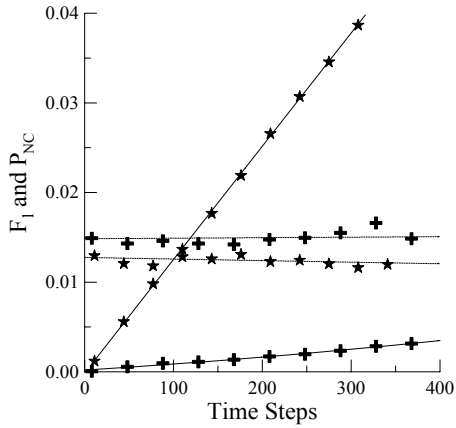

Figure 7. Fidelity $F_1$ and $P_{NC}$ against time steps for $\varepsilon = 1.25\ 10^{-4}$, $\gamma = 1.39\ 10^{-4}$ (C=0.9). Full lines $P_{NC}$, dashed lines $F_1$. Symbols: ✚ Steane parallelized ancilla (4); ✶ non-fault-tolerant ancilla (1).

The effect of the $\Delta t$ (number of time steps between consecutive corrections) variation is not important when a fault-tolerant recovery is used; so long as its effect when ancilla (1) is considered cannot be neglected especially for a small enough $\varepsilon$ and $\gamma$. Figure 6 shows the $\Delta t$ effect on $P_{NC}$ for the non-fault-tolerant ancilla (1). The $P_{NC}$ minimum gets bigger $\Delta t$ as the error decreases, with the optimum value at about 70-80 time steps. A non-fault-tolerant error correcting network only is advantageous when the error rate is small enough and the correction frequency is optimized. But even in this optimal case, a full fault-tolerant network such as the Shor's five qubit ancilla provides better results as figure 7 shows. Error accumulation (for the case $\varepsilon=1.25\ 10^{-4}$, C=0.9) is completely disastrous in the case of ancilla (1).

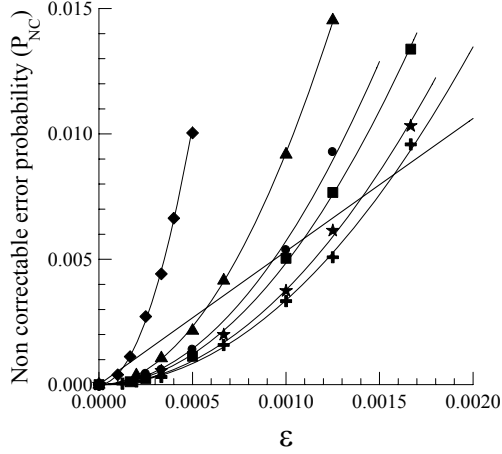

Figure 8. Non-correctable error probability ($P_{NC}$) against ε for different C values for a perfect $|0_L\rangle$ qubit sent through a noisy depolarizing quantum channel affected by ε evolution error and corrected by means of the ancilla (4). The gates have γ error probability. Symbols: ♦ C=0.1, fitting constant D=41143.425; ▲ C=0.3, D=9270.878; ● C=0.5, D=5729.543; ■ C=0.6, D=4855.088; ★ C=0.8, D=3778.015 and ✚ C=1, D=3368.333.

Finally, the objective will be to preserve a qubit stored in the memory of a quantum computer when it is affected by an ε evolution error probability. A noiseless encoded $|0_L\rangle$ qubit is sent through a noisy depolarizing quantum channel affected by an ε evolution error and corrected by means of a noisy recovering method affected by ε and γ error probabilities. Only the best ancilla (4) will be used to provide a memory error threshold curve $\varepsilon_{th}(C)$. Several estimations for the value of this threshold have been published [10,11] using different error models and correction schemes. Gottesman and Preskill [12], by means of the same [[7,1,3]] code and concatenation, estimate an error threshold rate of about $10^{-5}$ per time step, as memory error dominates. Following a method closer to the present one, Zalka [13] estimates the memory error threshold ε and the gate error threshold γ as one higher order of magnitude ($10^{-3}$) for the memory error. In an extensive treatment using different encoding and recovery strategies, Steane [14] finds a threshold value for quantum computation about $10^{-3}$. Recently Reichardt [15] provides a computation threshold of 9 $10^{-3}$, using the [[7,1,3]] quantum code and the depolarizing error model but without memory errors. The threshold obtained in the present work has a similar value.

In the non-encoded qubit case, the non-correctable error probability after t time steps is $P_{NE}(\varepsilon,t) = 1-(1-2\varepsilon/3)^t$. When fault-tolerant correction is used, the non-correctable errors (of weight two or more) do not accumulate very quickly, and the error probability will behave as $P_{NC} \sim O(\varepsilon^2,\gamma^2)$ (see figure 8). When the encoded qubit reaches the receiver, the information recovery process will be largely successful if the final decoding step can be performed without any error. The receiver fidelity will behave as 1-$O(\varepsilon^2,\gamma^2)$. So, for small enough $\varepsilon = C\gamma$, there must be an $\varepsilon$-region in which the threshold condition $P_{NC} \leq P_{NE}$ (for each C) is fulfilled. In the case of an encoded qubit, the first error correction is carried out at t = 8 time steps, so in figure 8 we represent the probabilities $P_{NE}(\varepsilon,t=8)$ and $P_{NC}(\varepsilon,C,t=8)$. The curves $P_{NC}(\varepsilon,C,t=8)$ are satisfactorily fitted to a quadratic polynomial $D(C)\ \varepsilon^2$, reflecting the fault-tolerance of the method. The crossing points between the curves $P_{NE}(\varepsilon,t=8)$ and $P_{NC}(\varepsilon,C,t=8)$, provide the error probability threshold $\varepsilon_{th}(C) \sim 16/3D$ which are represented in figure 9. The region under the curve $\varepsilon_{th}(C)$ is where the non-correctable error probability for the encoded qubit is smaller than for the non-encoded qubit. Given a noisy channel with a memory error probability $\varepsilon$, the $\varepsilon_{th}(C)$ curve give the gate error probability $0 < \gamma = \varepsilon/C$ (for the gates used in the recovery network) to produce a higher quality qubit state than without encoding. Note $C \leq 1$, because the gate implementation could include several time steps affected by an $\varepsilon$ memory error and $\gamma \geq \varepsilon$. The case C = 1 describes a gate carried out in one time step without intrinsic $\gamma$ error, and could be considered as a possible error threshold.

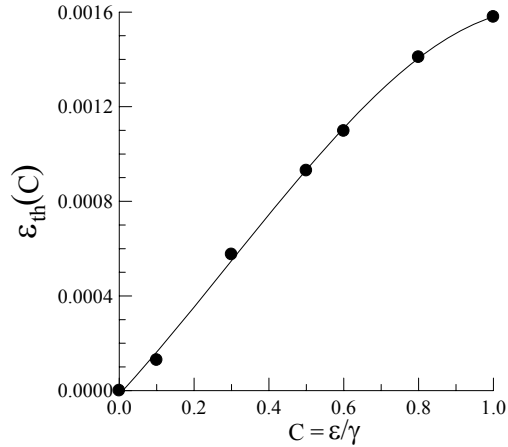

Figure 9. Memory error threshold ($\varepsilon_{th}(C) = 16/3D$) versus $C = \varepsilon/\gamma$ for the correction using the ancilla (4).

## 5 Conclusions

The present results show the possibility of using a simple non-fault-tolerant error correction if the error rate is small enough, the recovery frequency is optimized and the computation is not very long. If these conditions are not fulfilled, a full fault-tolerant error correction will be needed. Among the fault-tolerant ancillas, the (4), having a simpler and most parallelized implementation circuit, is the best of those considered in the paper and provides a memory error threshold of $\varepsilon_{th}(C=1) = 1.6\ 10^{-3}$. Its meaning is: if $\varepsilon < \varepsilon_{th}(C=1)$, there exists a set of noisy gates with $0 < \gamma = \varepsilon/C$ error probability, capable of stabilising an encoded qubit longer than an unencoded qubit. The agreement between the present threshold and the values obtained by some other authors with different error models, point us to its correctness.

**Acknowledgements**

The author thanks Mark Hallett for reviewing the paper.

**References**

1. P. W. Shor, Phys. Rev. A **52**, R2493 (1995). C. H. Bennett, D. DiVincenzo, J.A. Smolin, and W.K. Wootters, Phys. Rev. A **54**, 3824 (1996) (quant-ph/9604024).
2. P. Shor, in *Proceedings of the 37th Symp. on Found. of Comp. Science*, Los Alamitos, California, 56, (1996) (quant-ph/9605011). J. Preskill, in *Introduction to quantum computation* (World Scientific, Singapore, 2000), Chap. 8, p.213-269 (quant-ph/9712048 (1997).
3. R. Laflamme C. Miquel, J.P. Paz, and W.H. Zurek, Phys. Rev. Lett. **77**, 198 (1996).
4. A. R. Calderbank and P. Shor, Phys. Rev. A 54, 1098 (1996).
5. E. Knill, R. Laflamme and W. H. Zurek, Proc. R. Soc. London A **454**, 365 (1998) (quant-ph/9702058).
6. P. J. Salas and A.L. Sanz, Phys. Rev. A **66**, 022302 (2002) (quant-ph/0207068).
7. P. J. Salas and A.L. Sanz, Phys. Rev. A **69**, 052322 (2004).
8. M. Luscher, Comp. Phys. Com., **79**, 100 (1994).
9. F. James, Comp. Phys. Com., **79**, 111 (1994).
10. D. Aharonov and M. Ben-Or, *Proc. of 29th Annual ACM Symposium on Theory of Computing (STOC)* (El Paso, Texas, 1997) (quant-ph/9611025). E. Knill, R. Laflamme and W. Zurek, Technical Report LAUR-96-2199, Los Alamos National Laboratory (1996) (quant-ph/9610011).
11. D. Gottesman, Ph.D. thesis, California Institute of Technology (1997) (quant-ph/9705052).
12. J. Preskill, in *Introduction to quantum computation*, 3rd edition, edited by H. Lo, S. Popescu and T. Spiller (World Scientific, Singapore, 2000) (quant-ph/9712048). D. Gottesman, Phys. Rev. A 57, 127 (1998).
13. C. Zalka, technical report quant-ph 9612028.
14. A. M. Steane, technical report quant-ph/0207119.
15. B.W. Reichardt, technical report quant-ph/0406025.